\documentclass[twocolumn,epjc3]{svjour3}

\RequirePackage[T1]{fontenc}
\RequirePackage{fix-cm}
\RequirePackage[colorlinks,citecolor=blue,urlcolor=blue,linkcolor=blue]{hyperref}
\RequirePackage{siunitx}
\RequirePackage{booktabs}
\RequirePackage{upgreek}
\RequirePackage[numbers,sort&compress]{natbib}
\RequirePackage{graphbox}
\RequirePackage{amsmath,amssymb, gensymb}
\RequirePackage{url}
\RequirePackage{xspace}
\RequirePackage{textcomp}
\RequirePackage{microtype} 
\RequirePackage{float}
\RequirePackage[switch, modulo]{lineno}
\RequirePackage{mathptmx}      

\modulolinenumbers[2]

\usepackage{siunitx}
\usepackage{xspace}

\newcommand{\NEW}{NEXT-White\xspace}
\newcommand{\NEXT}{NEXT-100\xspace}

\newcommand{\Fig}{Figure}

\newcommand{\Tab}{Table}

\newcommand{\bbonu}{\ensuremath{\beta\beta0\nu}\xspace}

\newcommand{\Qbb}{\ensuremath{Q_{\beta\beta}}\xspace}

\newcommand{\Tonu}{\ensuremath{T_{1/2}^{0\nu}}\xspace}

\newcommand{\KR}{\ensuremath{^{83\textrm{m}}\mathrm{Kr}}\xspace}
\newcommand{\RB}{\ensuremath{^{83}\mathrm{Rb}}\xspace}

\newcommand{\XE}{\ensuremath{{}^{136}\rm Xe}\xspace}
\newcommand{\GE}{\ensuremath{{}^{76}\rm Ge\xspace}}

\DeclareSIUnit\c{\mbox{$c$}}
\DeclareSIUnit\magn{\mbox{$\times$}}
\DeclareSIUnit\min{min}
\DeclareSIUnit\week{week}
\DeclareSIUnit\year{yr}
\DeclareSIUnit\years{years}
\DeclareSIUnit\yr{yr}
\DeclareSIUnit\standard{std}
\DeclareSIUnit\str{sr}
\DeclareSIUnit\ppm{ppm}
\DeclareSIUnit\ppb{ppb}
\DeclareSIUnit\ppt{ppt}
\DeclareSIUnit\pe{PE}
\DeclareSIUnit\spe{SPE}
\DeclareSIUnit\ev{events}
\DeclareSIUnit\ct{counts}
\DeclareSIUnit\neutron{\mbox{$n$}}
\DeclareSIUnit\smp{samples}
\DeclareSIUnit\Sample{S}
\DeclareSIUnit\ch{ch}
\DeclareSIUnit\hit{hit}
\DeclareSIUnit\hits{hits}
\DeclareSIUnit\bin{(\mbox{5-PE}~bin)}
\DeclareSIUnit\sgm{\mbox{$\sigma$}}
\DeclareSIUnit\rms{RMS}
\DeclareSIUnit\keVr{\mbox{keV$_{\rm nr}$}}
\DeclareSIUnit\keVee{\mbox{keV$_{e{\rm e}}$}}
\DeclareSIUnit\ph{photon}
\DeclareSIUnit\pes{pes}
\DeclareSIUnit\el{electrons}
\DeclareSIUnit\pm{PMT}
\DeclareSIUnit\inch{"}
\DeclareSIUnit\bit{bit}
\DeclareSIUnit\sample{samples}
\DeclareSIUnit\barn{barn}
\DeclareSIUnit\bara{bar}
\DeclareSIUnit\barg{barg}
\DeclareSIUnit\mlardepth{\mbox(meter~of~\LAr~depth)}
\DeclareSIUnit\Curie{Ci}
\DeclareSIUnit\psi{psi}
\DeclareSIUnit\parsec{pc}
\DeclareSIUnit\liveday{\mbox{live-days}}
\DeclareSIUnit\days{\mbox{days}}
\DeclareSIUnit\day{\mbox{day}}
\DeclareSIUnit\miles{\mbox{miles}}
\DeclareSIUnit\degreeC{\mbox{$^{\circ}$C}}
\DeclareSIUnit\electron{\mbox{$e^-$}}
\DeclareSIUnit\Euro{\mbox{\euro}}
\DeclareSIUnit\cph{cph}
\DeclareSIUnit\neq{neq}
\DeclareSIUnit\unit{unit}
\DeclareSIUnit\byte{Byte}
\DeclareSIUnit\Bq{\becquerel}

\journalname{Eur. Phys. J. C}

\begin{document}


\title{First results of the \NEXT\ detector using {\boldmath\KR} decays}

\author{
 G. Mart\'inez-Lema\thanksref{gml, ca1, addr2} 
\and
 C. Herv\'es Carrete\thanksref{addr8} 
\and
 S. Torelli\thanksref{addr2} 
\and
 M. Cid Laso\thanksref{addr7,addr8} 
\and
 P. V\'azquez Cabaleiro\thanksref{addr2,addr8} 
\and
 B. Palmeiro\thanksref{addr15} 
\and
 J.A. Hernando~Morata\thanksref{ca2, addr8} 
\and
 J.J. G\'omez-Cadenas\thanksref{f41,addr2,addr17} 
\and
 C. Adams\thanksref{f2,addr1} 
\and
 H. Almaz\'an\thanksref{addr2}
% ...
}

\thankstext{gml}{Now at Instituto de F\'isica Corpuscular, Spain.}
\thankstext{f2}{Now at NVIDIA.}
\thankstext{f26}{Deceased.}
\thankstext{f35}{Now at Weizmann Institute of Science, Israel.}
\thankstext{f37}{On leave.}
\thankstext{f41}{NEXT Spokesperson.}
\thankstext{f89}{Now at University of North Carolina, USA.}
\thankstext{f90}{Now at Los Alamos National Laboratory, USA.}

\thankstext{ca1}{e-mail: gonzalo.martinez.lema@ific.uv.es}
\thankstext{ca2}{e-mail: jose.hernando@usc.es}

\institute{
Argonne National Laboratory, Argonne, IL 60439, USA \label{addr1} 
\and
 Donostia International Physics Center, BERC Basque Excellence Research Centre, Manuel de Lardizabal 4, San Sebasti\'an / Donostia, E-20018, Spain \label{addr2} 
% ...
\and
 Instituto de F\'isica Corpuscular (IFIC), CSIC \& Universitat de Val\`encia, Calle Catedr\'atico Jos\'e Beltr\'an, 2, Paterna, E-46980, Spain \label{addr7} 
\and
 Instituto Gallego de F\'isica de Altas Energ\'ias, Univ.\ de Santiago de Compostela, Campus sur, R\'ua Xos\'e Mar\'ia Su\'arez N\'u\~nez, s/n, Santiago de Compostela, E-15782, Spain \label{addr8} 
% ...
\and
 Department of Physics and Astronomy, Manchester University, Manchester. M13 9PL, United Kingdom \label{addr15} 
% ...
\and
 Ikerbasque (Basque Foundation for Science), Bilbao, E-48009, Spain \label{addr17} 
% ...
}

\date{Received: date / Accepted: date}

\maketitle
\flushbottom

\begin{abstract}
The NEXT collaboration is investigating the double beta decay of \XE\ using high-pressure gas electroluminescent time projection chambers, which provide excellent energy resolution together with a robust topological signature. Operating at the Laboratorio Subterráneo de Canfranc (LSC) and building on the success of the NEXT-White detector, the NEXT-100 apparatus began commissioning in May 2024 and started operation with xenon at a pressure of 4 bar in October 2024. 

We report here the first results obtained with \NEXT\ using low-energy calibration data from \KR\ decays, which allow mapping of the detector response in the active volume and monitoring of its stability over time. After homogenizing the light response, we achieve an energy resolution of 4.37\% FWHM at 41.5~keV for \KR\ point-like energy deposits contained in a radius of 425~mm. In a fiducial region representing the operating conditions of NEXT-100 at 10 bar we obtain an improved energy resolution of 4.16\% FWHM. These results are in good agreement with that obtained in \NEW, and an $E^{-1/2}$ extrapolation to \Qbb\ yields an energy resolution close to 0.5\% FWHM, well below the 1\% FWHM design target.


\end{abstract}

\section{Introduction}
\label{sec:introduction}

Neutrinoless double beta decay (\bbonu) is a putative rare nuclear decay that violates lepton number conservation. An observation would unequivocally establish the Majorana nature of the neutrino. Currently, the most stringent constraints on the half-life of this process (\Tonu) are set by the KamLAND-Zen experiment at $2.23\times 10^{26}$~yr (90\% C.L.) for \XE~\cite{KamLAND-Zen:23}, and by the LEGEND-200 experiment at $1.9\times 10^{26}$~yr (90\% C.L.) for \GE~\cite{25tk-nctn}.

NEXT is a \bbonu experiment searching for this decay in gaseous \XE using a High Pressure Time Projection Chamber with ElectroLuminescent amplification (HPXeTPC-EL). After a series of prototypes~\cite{DEMO1,DEMO2,DEMO3,DEMO4}, the NEXT Collaboration operated \NEW~\cite{NEWDetector}, a mid-scale HPXeTPC-EL holding $\sim$5~kg of \XE-enriched xenon at 10 bar hosted at the Laboratorio Subterráneo de Canfranc (LSC). With \NEW, the Collaboration provided a direct estimation of sub-percent FWHM energy resolution at \Qbb~\cite{NEWEres1, NEWEres2}, demonstrated the topological discrimination between single- and double-electron tracks~\cite{Kekic2021, NEWTrack1, NEWTrack2}, measured the main backgrounds of the experiment~\cite{NEWBkg1, NEWBkg2}, obtained the first measurement of the half life of the two-neutrino mode of the decay using a novel technique~\cite{NEW2nubb}, and established a limit on \Tonu~\cite{NEW0nubb}.

Following the success of its predecessor, the Collaboration built NEXT-100, a ${\sim}10\times$ larger apparatus (by volume) that began operations in 2024 \cite{NEXT:2025yqw}. The detector was operated with depleted xenon at a pressure of 4 bar, while its nominal operation at high pressure (13.5 bar) is foreseen to begin in early 2026.

Here we report the first results obtained during the calibration run with low-energy events in \NEXT. Following a procedure similar to the one employed in \NEW~\cite{NEWKR}, \KR decays were used to characterize the detector in the active volume. This enabled the mapping of the light response as a function of $(x, y, z)$. The variations in $(x,y)$ arise from a combination of the dependence of the light collection efficiency due to variations on the effective solid angle coverage, and from inhomogeneities in the electroluminescence (EL) field. Variations along the longitudinal coordinate ($z$) originate mainly from electron attachment. This method was also used to assess the performance of the detector at low energies with point-like events, and to monitor its stability over time, allowing for a continuous calibration of the detector during data-taking campaigns. These \KR calibrations allow for the homogenization of the detector response, which plays a pivotal role in the overall performance of the detector at high energies.

The article is structured as follows. In Section~\ref{sec:NEXT100_detector}, we provide a brief description of the \NEXT detector. In Section~\ref{sec:calibration_procedure}, we describe the data-taking campaign and the \KR\ calibration procedure, and present the analysis of the detector response in the active volume. In Section~\ref{sec:stability}, we analyze the detector stability over the calibration run. In Section~\ref{sec:performance}, we evaluate the energy resolution obtained for \KR deposits and its variation over the active volume. The results and conclusions are summarized in Section~\ref{sec:conclusions}.

\section{The \NEXT detector}
\label{sec:NEXT100_detector}

\begin{figure}
    \centering
    \includegraphics[width=\columnwidth]{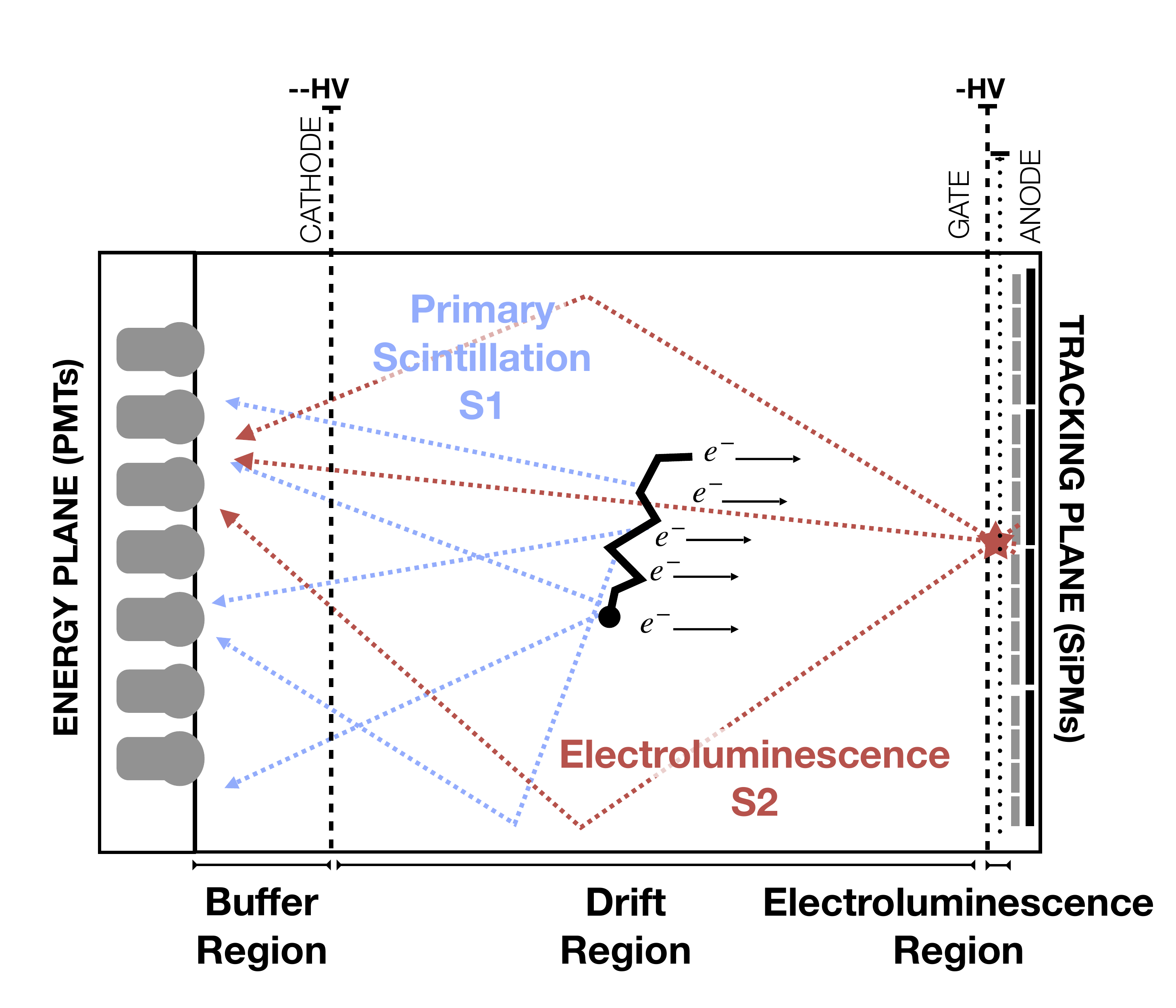}
    \caption{Schematic drawing of the \NEXT detector. Reproduced from \cite{NEXT:2025yqw}.}
    \label{fig:detector-scheme}
\end{figure}

\NEXT is a cylindrical HPXeTPC-EL instrumented with photosensor planes at each end of the chamber. On one end, the \emph{energy plane} hosts photomultiplier tubes (PMTs), used to measure energy and the start time of the event. On the opposite side of the detector, the \emph{tracking plane} features an array of silicon photomultipliers (SiPMs), used to extract topological information from the events. With an inner diameter of 98.3~cm, the detector volume is divided into three regions with different electric fields by three stainless steel meshes with hexagonal openings acting as cathode, gate and anode. The \emph{drift region}, bounded by the cathode and the gate, is 118.7~cm long. The \emph{EL region} adjacent to the drift region is a 0.97~cm long gap defined by the gate and the anode. The tracking plane is located 15~mm away from the anode. The space between the energy plane and the cathode defines the \emph{buffer region}, a 24~cm long gap that prevents sparks between the cathode and the energy plane and suppresses the production of unwanted electroluminescence in this region.
A scheme of the \NEXT detector is shown in \Fig~\ref{fig:detector-scheme}; a detailed description of the NEXT-100 detector can be found in~\cite{NEXT:2025yqw}.

The energy plane is designed to hold 60 PMTs (Hamamatsu, model R11410-10), of which 48 were fully operational during the data-taking period. 7 PMTs were not installed as some surrounding components did not pass the quality test. The remaining 5 PMTs yielded a bad signal, likely due to electronic issues, and were turned off. For the upcoming high-pressure run, all 60 PMTs are expected to be operative.
The tracking plane comprises 3584 SiPMs (Hamamatsu S13372-1350TE $1.3\times1.3$~mm$^2$), with just 4 of them offline. Each PMT is housed in an individual vacuum-insulated capsule, optically coupled to the active volume through a sapphire window. The PMTs are arranged in a hexagonal (honeycomb) pattern, while the SiPMs are mounted on modular boards each holding a matrix of $8\times8$ sensors in an square pattern, with a 15.55~mm pitch between adjacent sensors. The SiPM array extends beyond the active area of the detector, avoiding dead regions.


\subsection{Principle of operation}

An ionizing interaction within the active volume produces two distinct signals: an initial prompt scintillation flash (S1), detected by the PMTs on the energy plane, and a delayed EL signal (S2), generated by ionization electrons after they reach the EL region. These electrons drift toward the gate thanks to a moderate drift field (of the order of 100~V/cm) at a velocity of the order of 1~mm/$\upmu$s and enter the EL region, where they are accelerated by an intense electric field (tens of kV/cm), producing a second flash of light. The $\sim$172~nm photons emitted by xenon (both S1 and S2) are wavelength-shifted to the visible range by a thin layer of tetraphenyl butadiene (TPB) vacuum-deposited on the inner surfaces of the detector, including the SiPM boards, sapphire windows, and the polytetrafluoroethylene (PTFE) panels lining the walls of the TPC. The resulting blue light is detected by both planes. In the tracking plane, the light pattern is focused near the production point, allowing for the reconstruction of the charge distribution in the traverse plane $(x, y)$. On the other hand, the light detected by the energy plane is more diffuse, which allows to integrate a larger fraction of the emitted light, providing an accurate measurement of the deposited energy. The combination of the $(x, y)$ position provided by the SiPMs and the drift time, defined as the delay between the S1 and S2 signals, enables full 3D reconstruction of the ionization track.

The two key features that allow \NEXT to effectively suppress background events are: (1) its excellent energy resolution, expected to be better than 1\% FWHM at \Qbb, and (2) its capability for topological discrimination between signal-like events (two-electron tracks) and background events (single-electron tracks). Signal events typically exhibit two energy depositions, or \emph{blobs}, located at the ends of the track, corresponding to the Bragg peaks of each electron. In contrast, background single-electron tracks produce only one such blob. This distinctive topological signature serves as a powerful handle for background rejection.
\section{{\boldmath\KR} calibration}
\label{sec:calibration_procedure}

We calibrated the \NEXT detector using \KR decays, following a similar methodology as applied previously to the \NEW detector, described in detail in~\cite{NEWKR}. \KR is generated via the decay of the radioactive isotope \RB, which has a half-life of 86.2~days. The \RB\ source, embedded in small 1~mm-diameter zeolite balls~\cite{KrSource}, is housed within a dedicated branch of the gas circulation system. \KR isotopes emanate from the zeolite and are continuously injected into the xenon gas stream, where they mix homogeneously with the xenon and diffuse into the active volume of the detector \cite{NEXT:2025yqw}.

\KR decays via two consecutive internal conversion processes. The first, with a half-life of 1.83~hours, dominates the decay dynamics; the second, much faster, occurs on a timescale of 154.4~ns. The decay deposits a total energy of 41.5~keV in a very short spatial range, allowing to treat it as point-like and providing an ideal mono-energetic calibration source to map the detector response.

\subsection{Operational conditions}

The \KR datasets analyzed in this study were acquired during June - September 2025. The main operational parameters during data taking are summarized in \Tab~\ref{tab:operation}.

\begin{table}[h]
\renewcommand{\arraystretch}{1.1}
\centering
\begin{tabular}{@{} l c r @{}} 
 \toprule
 Pressure &\ & 4~bar \\
 Temperature &\ & 22.5$^\circ$C \\
 Cathode voltage &\ & 23~kV \\
 Gate voltage &\ & 8.8~kV \\
 Drift field &\ & 120~V/cm \\
 Reduced EL field &\ & 2.27 ~kV/cm/bar \\
 \bottomrule
\end{tabular}
\caption{Operational parameters during data taking.}
\label{tab:operation}
\end{table}

\subsection{Data acquisition and processing}
\label{sec:reco}

\begin{figure*}
    \centering
    \includegraphics[width=0.9\textwidth]{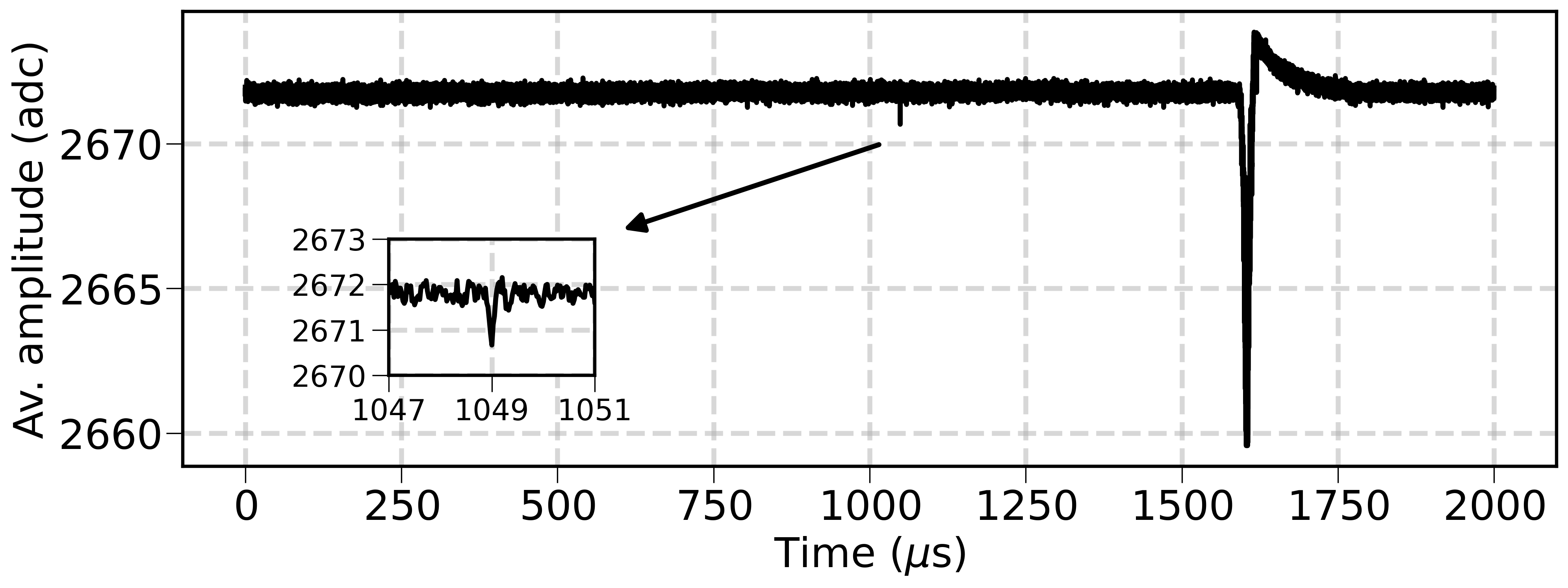}
    \caption{Averaged PMT waveform containing a \KR event candidate. The small first peak at $t\approx$1050~$\upmu$s corresponds to the S1 signal; a zoomed-in view in shown in the inset axis. In this example, the S1 amplitude is about 3 times larger than the standard deviation of the electronic noise. The second, larger peak at $t\approx$1600~$\upmu$s corresponds to the S2 signal.}
    \label{fig:kr_wf}
\end{figure*}

The DAQ system in NEXT operates in dual-trigger mode, with one trigger typically focusing on low-energy events (\KR-like), and the other on high-energy ones. This enables simultaneous data-taking for both \KR calibration and physics data.
\KR events are triggered when the signal in a central PMT exceeds a predefined threshold of 2 ADC counts for a time span of [0.5, 20]~$\upmu$s, allowing for a 400-$\upmu$s-long drop below threshold. The pulse height must be below 100 ADC counts and have an integrated charge in the range (1000, 15000) ADC counts. These conditions map roughly to the energy range (5, 100) keV and is designed to search for S2-like pulses. PMT waveforms are digitized at 25~ns sampling intervals over a buffer window of 2000~$\upmu$s. An example of a PMT-averaged waveform containing a \KR event candidate is shown in \Fig~\ref{fig:kr_wf}. The PMT front-end electronics produces a bipolar signal that is processed offline using a Baseline Restoration (BLR) algorithm to retrieve the unipolar signal produced by the PMT~\cite{PMTelectronics}. The SiPM waveforms, sampled at 1~$\upmu$s intervals, are zero-suppressed in real time by the DAQ system to reduce the data throughput.

\begin{figure}
    \centering
    \includegraphics[width=0.91\columnwidth]{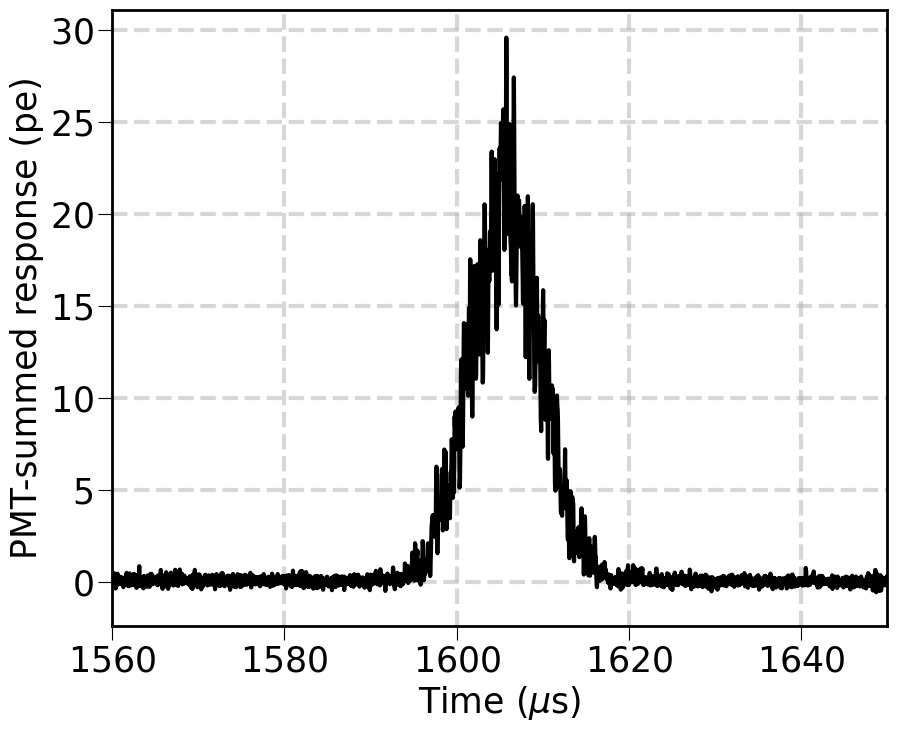}
    \includegraphics[width=1.00\columnwidth]{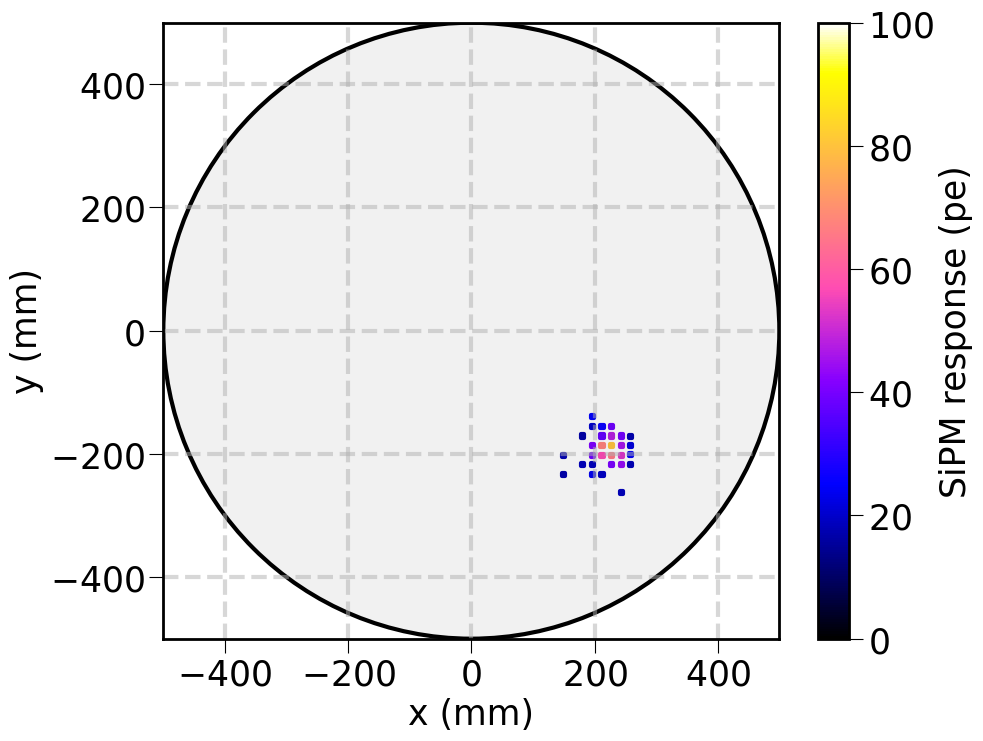}
    \caption{Top: calibrated PMT-summed response of a \KR event candidate zoomed in on the S2 pulse. Bottom: time-integrated SiPM response of a \KR event candidate.}
    \label{fig:kr_s2}
\end{figure}

The gain and pedestal noise of the PMTs are monitored weekly using dedicated sensor calibration runs with an array of blue LEDs installed on the tracking plane. Similarly, the SiPMs gain and noise are measured using a complementary LED system located on the energy plane. The SiPM and (BLR-processed) PMT waveforms are converted from ADC to photoelectrons (pe) using these calibration constants. Once calibrated, the PMT waveforms are summed to form a global waveform for each event. The top panel of \Fig~\ref{fig:kr_s2} shows a BLR-processed, PMT-summed waveform of a \KR event candidate, centered around the S2 pulse, which has a Gaussian shape.

Signal identification is performed on the summed PMT waveform. These waveforms are scanned twice to look for S1 and S2 signals independently. S1s are searched up to the trigger time, while S2s are searched over the full waveform. \KR S1 and S2 pulses are vastly different in terms of their width, height and amplitude: S2s are wide and high pulses, while S1s are narrow and low-amplitude peaks. Thus, a signal is classified as S2 if the waveform deviates from the baseline more than 0.5 pe for a time span greater than 4 $\upmu$s. In contrast, \KR S1 signals can be masked by the intrinsic PMT noise and periodic oscillations of the baseline. Therefore, the S1 peak search is performed by looking for deviations of the waveform from a local estimation of the baseline. This local estimation is obtained using a moving average window of 2.5 $\upmu$s, which accounts for small-scale fluctuations in the baseline. A peak is classified as S1 if the waveform deviates at least 0.1 pe from the local baseline for a time span between 0.125~$\upmu$s and 1~$\upmu$s.

Each peak reconstructed is further processed to obtain some basic information about the pulse, such as width and integrated PMT signal. The reconstructed \KR events display a mean S1 integrated signal of $\approx$9.6 pe, while the S2 signal yields around 8500 pe. The S2 SiPM information is used to compute the center of gravity (barycenter) of the tracking plane response, providing an $(x, y)$ reconstruction of the S2 pulse. Only SiPMs with a response above 10 pe are considered. The bottom panel of \Fig~\ref{fig:kr_s2} shows a typical SiPM response of a \KR event candidate. By comparing the S1 and S2 times, we obtain the drift time, which is used to determine the $z$ position, achieving a full 3D reconstruction of the energy deposit.
For each event, all possible S1–S2 peak pairs are generated.


\subsection{Data selection}
\label{sec:selection}

\begin{figure}
    \centering
    \includegraphics[width=\columnwidth]{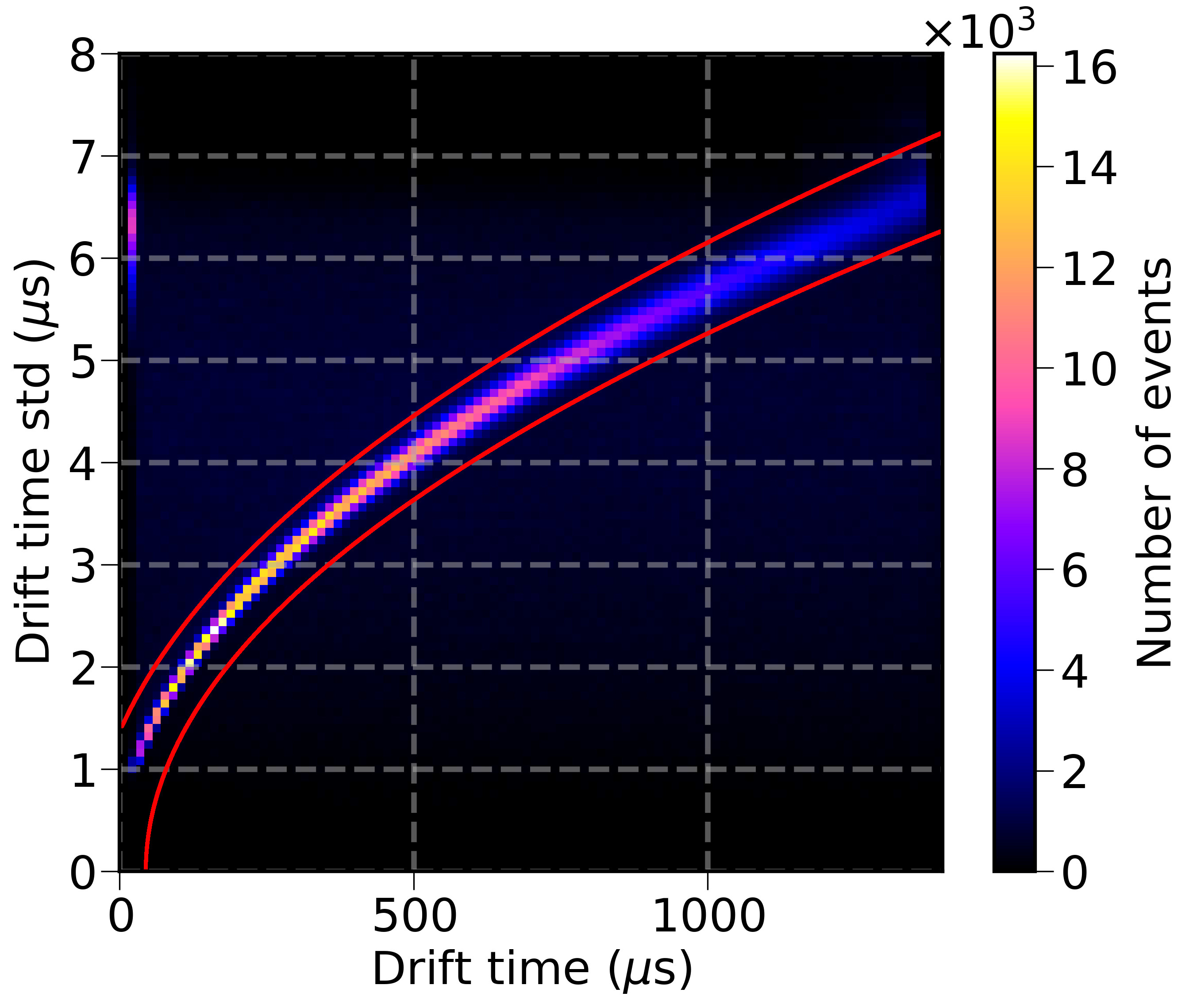}
    \caption{S2 time standard deviation as a function of drift time. Events with a single S1 and S2 falling within the band defined by the two lines are selected as \KR\ candidates.}
    \label{fig:selection}
\end{figure}

In order to minimize reconstruction bias, the peak-finding parameters described above are highly relaxed. This is particularly critical in the case of S1 signals from \KR decays, which are at the limit of our reconstruction capabilities. To disregard S1 misidentification and associate the true S1 signal with the corresponding S2 peak, we exploit the correlation between the drift time and the longitudinal dispersion \cite{NEXT:2023jwt, NEWdif} of the S2 peak resulting from electron diffusion during their drift toward the EL region.
\Fig~\ref{fig:selection} displays the standard deviation in drift time of the S2 pulse as a function of drift time for all S1-S2 pairs in a dataset. A clear band is observed, corresponding to physical \KR decays where the S1 is correctly linked to an S2 signal. In turn, the faint homogeneous background in this Figure corresponds to misidentified S1 peaks or erroneously associated S1-S2 pairs. The hotspot on the top-left corner corresponds to fluctuations in the raising edge of the S2 signal which are reconstructed as S1 signals.

We select S1-S2 peak pairs within the band defined by the red lines shown in \Fig~\ref{fig:selection}. After this band selection, the fraction of pairs with only one S1 signal is above 80\%. The fraction of events with one S2 signal is close to 100\%.
In order to increase the purity of the \KR sample, only events with a single S1-S2 peak pair are retained.
This criterion ensures a robust identification of the true S1 signal and, consequently, an accurate measurement of the drift time.

\subsection{Estimation of the drift velocity}
\label{sec:driftv}

\begin{figure}
    \centering
    \includegraphics[width=\columnwidth]{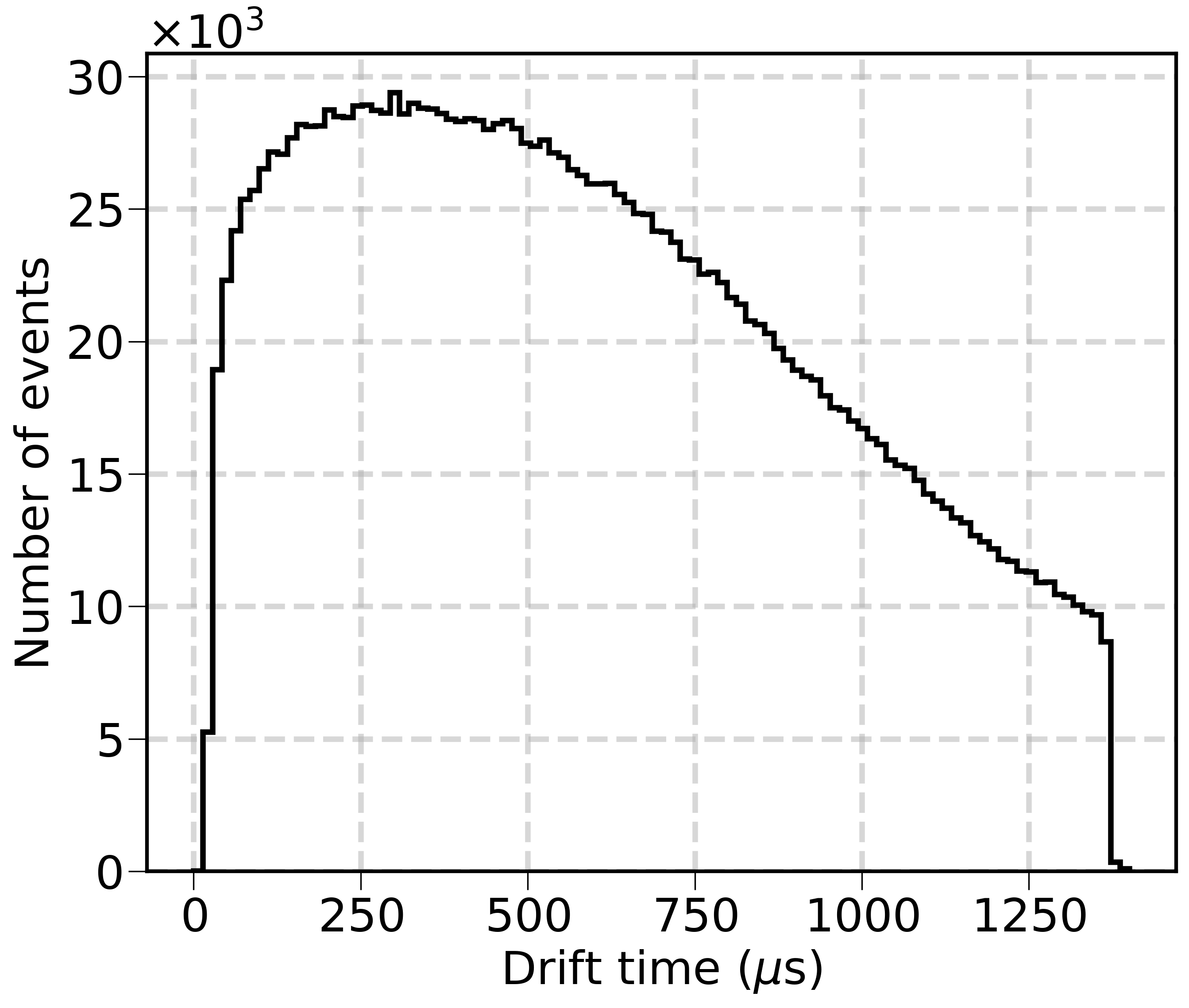}
    \includegraphics[width=\columnwidth]{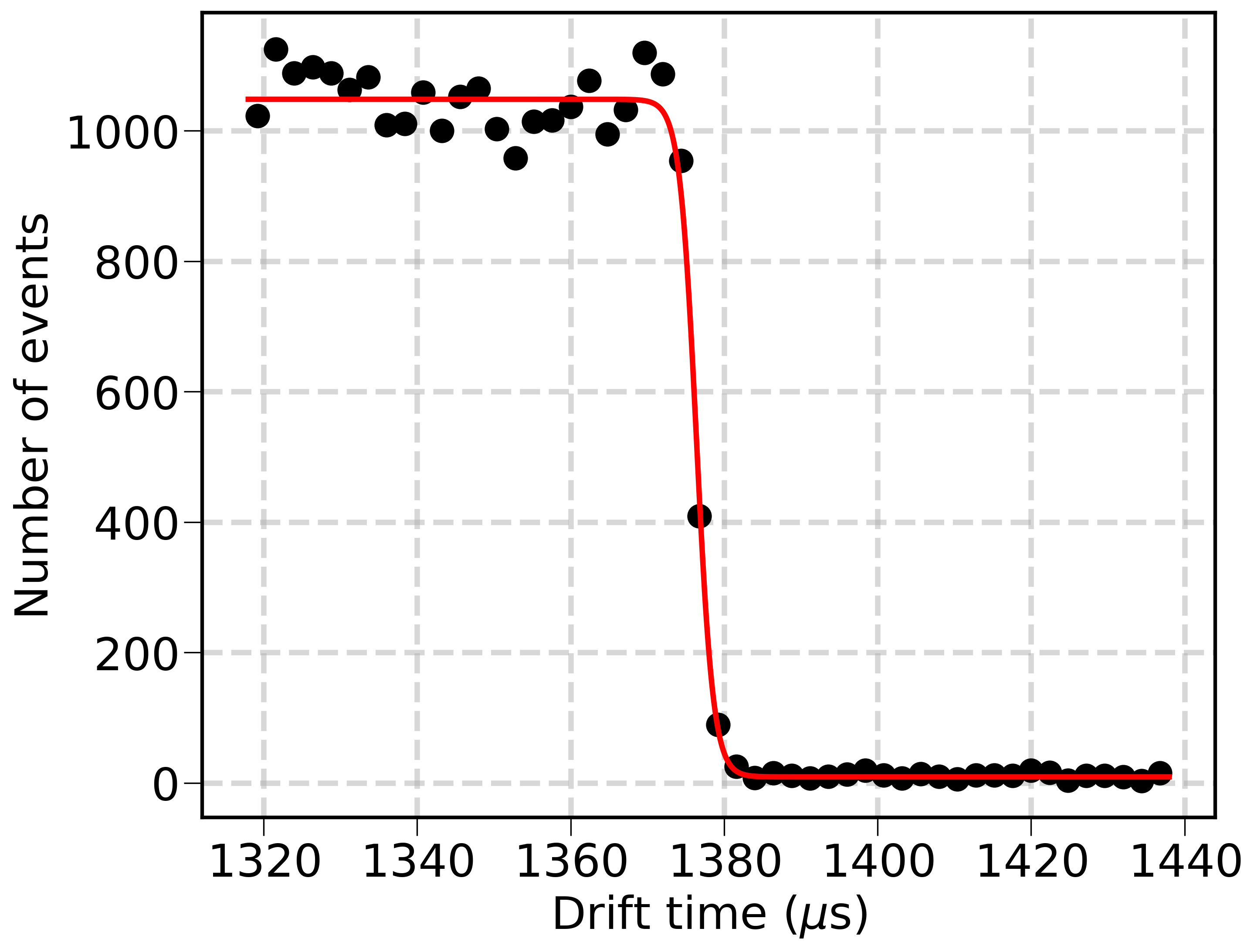}
    \caption{Drift time distribution of selected events (top), and a zoomed-in view of the end of the drift-time interval (bottom), with a sigmoid fit used to extract the drift velocity.}
    \label{fig:drift_time}
\end{figure}

The distribution of drift times for the selected \KR events is shown in the top panel of \Fig~\ref{fig:drift_time}. We expect a uniform event population in the active volume. The fact that it is not entirely homogeneous can be attributed to two main causes: (1) a lower S1 detection and selection efficiency at short drift times due to the lower geometrical acceptance, and\linebreak (2) a lower trigger and reconstruction efficiency for events at long drift times due to the spread of signal due to diffusion. The observed detection and selection inefficiencies are consistent with our observations in the Monte Carlo simulation and have also been observed in \NEW. The trigger and reconstruction effects have been confirmed by an independent analysis using a random trigger run.

The drift velocity in the detector is measured by analyzing the upper cutoff of the drift time distribution, which corresponds to depositions near the cathode. The distribution in this region is fitted to a sigmoid function to accurately determine the maximum drift time, $t_{d, \text{max}}$, defined as the inflection point of the sigmoid. Knowing the physical drift length of the chamber, $L_{\text{drift}} = (1187 \pm 2)$~mm~\cite{NEXT:2025yqw}, and the average time taken by an electron to traverse the EL gap, $t_{EL} = (1.783 \pm 0.028)~\upmu$s, the electron drift velocity is computed as

$$v_d = \frac{L_{\text{drift}}}{t_{d, \text{max}} - t_{EL}/2}.$$

\noindent
The factor $t_{EL}/2$ accounts for the fact that the drift time is computed from the maximum height of the S2 peak, which occurs at the center of the EL gap.
$t_{EL}$ was computed from the size of the EL gap, $d_{\text{EL}}= (9.70 \pm 0.15)$~mm \cite{NEXT:2025yqw}, and the drift velocity in the EL region computed from MAGBOLTZ \cite{MAGBOLTZ}. 
The bottom panel of \Fig~\ref{fig:drift_time} shows the distribution of events near the cathode and the corresponding fit to a sigmoid. We obtain $t_{d, max} \approx 1370.7~\upmu$s which results in a drift velocity $v_d=(0.867 \pm 0.001)$~mm/$\upmu$s. This result is in agreement to the one performed with alpha particles during the commissioning run \cite{NEXT:2025yqw}, and both numbers are $\approx$2.6\% higher than the MAGBOLTZ prediction for xenon at 4~bar under the applied electric field conditions:\linebreak$v_{d, \textrm{MB}}$ = (0.8455 $\pm$ 0.0072)~mm/$\upmu$s. Note, however, that other measurements with previous NEXT detectors were also consistently higher than the MAGBOLTZ prediction~\cite{DEMO_2013, Lorca_2014, NEWdif}. This could potentially be explained by an increase in the drift field in the vicinity of the cathode and gate meshes.


\subsection{Detector response}
\label{sec:detector_response}

\begin{figure*}
    \centering
    \includegraphics[width=\textwidth]{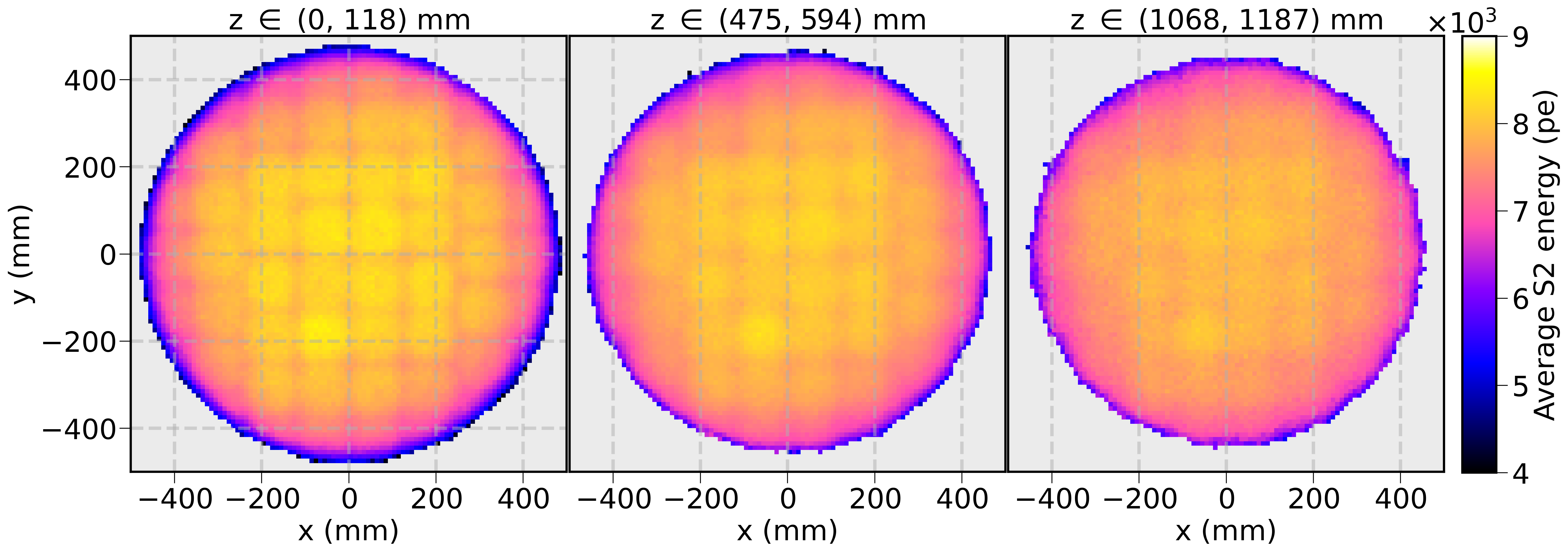}
    \caption{Average S2 signal for \KR events as a function of the $(x, y)$ position for different ranges of $z$. These maps can be interpreted as the S2 detector response in the full active volume.}
    \label{fig:E0_map}
\end{figure*}

The amount of light collected for a \KR event depends on its location on the $(x, y)$ plane due to geometrical variations in light collection efficiency across the EL plane, and on its drift time $t_d$ (or, equivalently, the coordinate $z=v_d\cdot t_d$) due to electron attachment effects. As demonstrated in \cite{NEXT:2025yqw}, the electron lifetime is much longer than the maximum drift time and therefore its associated correction is small. Furthermore, the measurement of the (x,y) response depends also on the drift coordinate due to electron diffusion. A pointlike energy deposition near the cathode produces an electron cloud with a dispersion of $\sim$30~mm in (x,y) and $\sim$7~mm in $z$. Therefore, the response contains a range of geometrical acceptance factors, and the events at longer drifts are susceptible to electron losses to the TPC walls, particularly at high radii.
To characterize these spatial dependencies, we compute the average S2 signal registered by the PMTs in voxels of $10\times10\times118$~mm$^3$ across the $(x, y,z)$ volume. \Fig~\ref{fig:E0_map} displays the average S2 energy as a function of $(x, y)$ for different ranges in $z$, which is effectively a map of the S2 detector response in the active volume.

In this map, we observe a clear radial dependence of the S2 light yield, while it is fairly symmetric in the azimuthal coordinate. Moreover, the response remains relatively uniform within a radius of 300~mm but gradually decreases beyond this region.
At the same time, the map features wide square regions of higher response interleaved with narrow regions of lower response. This pattern matches the boundaries of the SiPM boards, and is also visible in our Monte Carlo simulation. Hence, we attribute this effect to the difference in reflectivity between the TPB-coated surfaces of the SiPM boards and the gaps between them. The difference in response becomes less pronounced for longer drift times, consistent with the increased spatial spread of the electron cloud due to diffusion, which diminishes the contribution of these features.

These maps are produced on a run-by-run basis, each period typically lasting $\sim$24~h, with approximately $2 \cdot 10^6$ events selected for this process. Each voxel had a sample size of $\sim$30~events for a statistical uncertainty on the average energy of $\sim$0.3\%. As discussed in Section~\ref{sec:stability}, the stability of the maps allows us to combine measurements from consecutive runs, which further improves the uncertainty below 0.1\%. Therefore, the contribution of the energy corrections to the energy resolution is at most 0.25\% FWHM. 
\section{Detector stability}
\label{sec:stability}

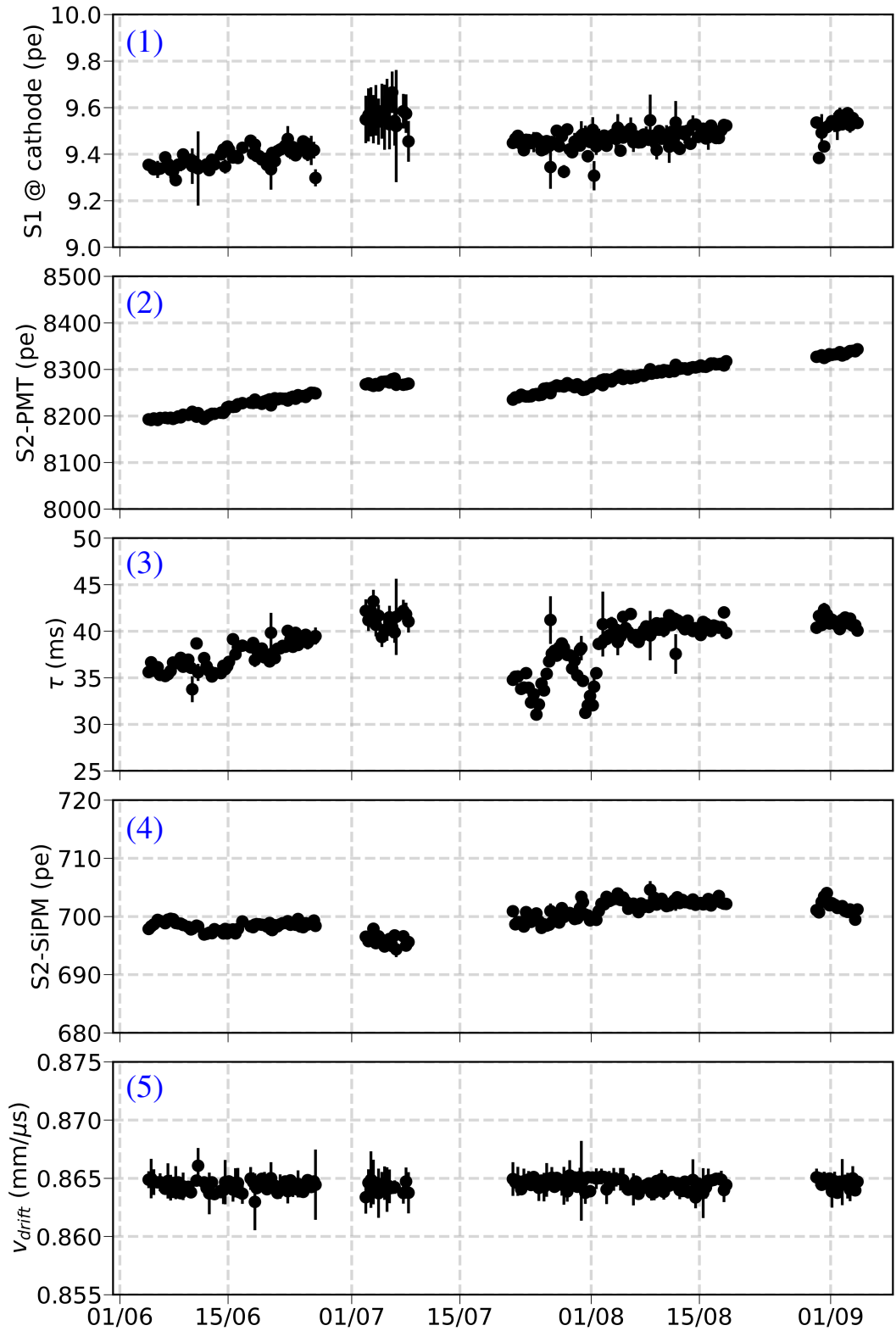
\begin{figure*}[p]
    \centering
    \includegraphics[width=0.8\textwidth]{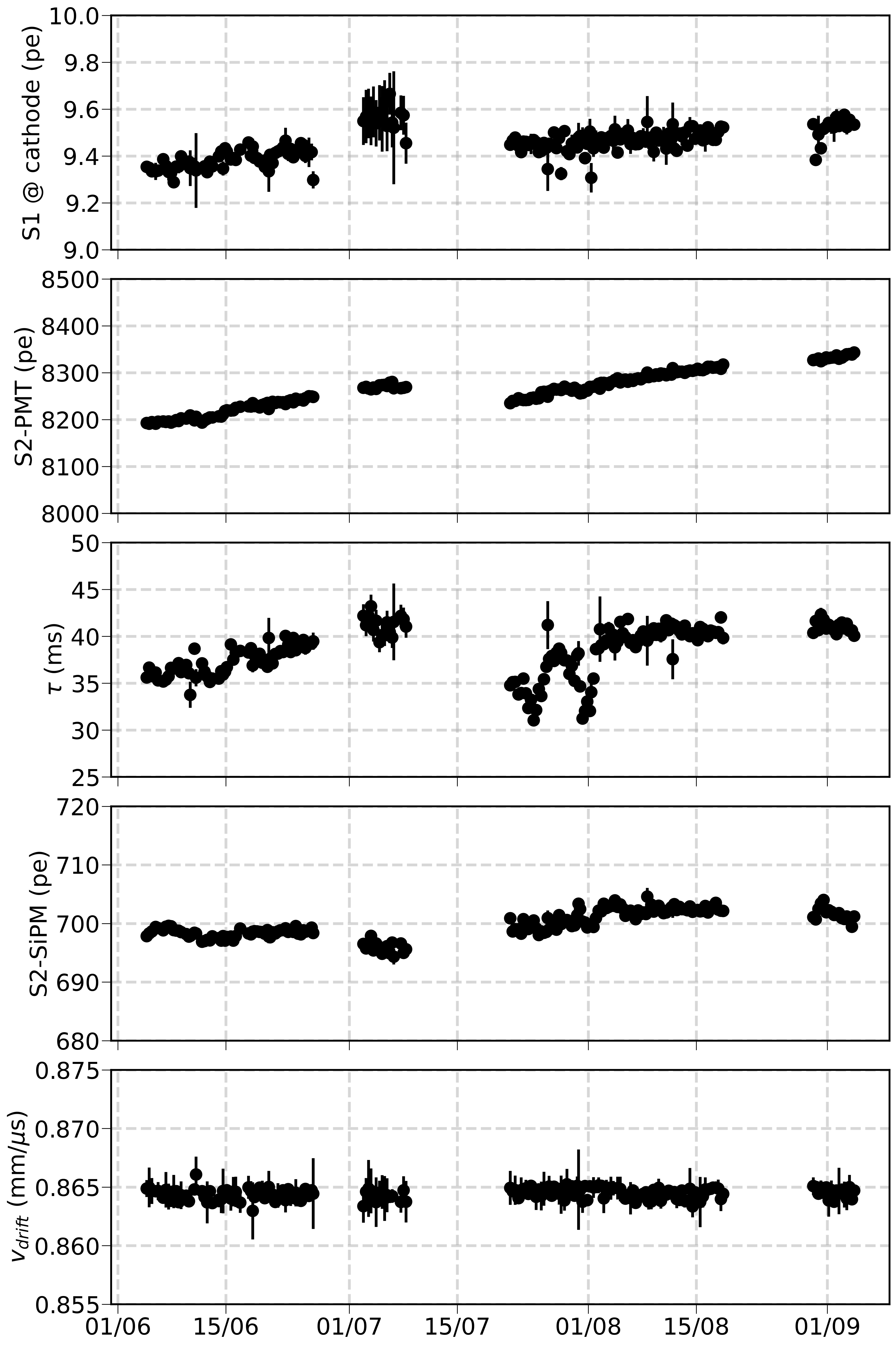}
    \begin{picture}(0,0)(335, -110)
        \put(0,  0){\makebox(0,0)[c]{{\Large \textcolor{blue}{(5)}}}}
        \put(0,115){\makebox(0,0)[c]{{\Large \textcolor{blue}{(4)}}}}
        \put(0,233){\makebox(0,0)[c]{{\Large \textcolor{blue}{(3)}}}}
        \put(0,349){\makebox(0,0)[c]{{\Large \textcolor{blue}{(2)}}}}
        \put(0,465){\makebox(0,0)[c]{{\Large \textcolor{blue}{(1)}}}}
    \end{picture}%
    \caption{Time evolution of several monitoring variables in the detector. From top to bottom: (1) average S1 signal for events near the cathode; (2) average \KR S2 signal after corrections; (3) electron lifetime; (4) integrated tracking-plane response for events near the anode; and (5) drift velocity.}
    \label{fig:evolution}
\end{figure*}

\KR calibrations were used to monitor \NEXT over long periods of time. The detector has demonstrated excellent operational stability during this initial data-taking period. The main quantities that are tracked are: (1) the integrated S1 signal, which does not depend on the configuration of the drift or EL fields; (2) the integrated S2 signal, which provides a measure of variations in EL amplification; (3) the electron lifetime, which serves as an indicator of the gas purity; (4) the integrated tracking plane light signal for events near the anode, to estimate the stability of the SiPM response; and (5) the drift velocity, which allows us to monitor the drift field. Here, the electron lifetime describes the process of electron attachment, which reduces the amount of electrons reaching the EL amplification region following an exponential law:

$$N_{EL} = N_0 \cdot e^{-t_d/\tau},$$

\noindent
where $N_0$ and $N_{EL}$ are the number of electrons produced and reaching the EL region, respectively, $t_d$ is the drift time, and $\tau$ is the electron lifetime.


The top panel of \Fig~\ref{fig:evolution} shows the evolution of the average S1 light yield, from \KR events near the cathode as a function of time. The second panel of the same figure shows the temporal evolution of the average S2 signal after corrections. Both S1 and S2 quantities exhibit a mild S2 variation of $\lesssim 2$\% over this three-month period. As both quantities exhibit the same trend and variation rate, we conclude it is unrelated to the EL amplification process. Moreover, since the SiPM response in the fourth panel (also related to the light yield) is rather flat, we attribute this effect to a slow drift in the PMT gain.

The third panel shows the evolution of the electron lifetime, $\tau$, as a function of time. The lifetime improved monotonically during the first half of this period. Subsequently, gas circulation was paused twice to insert a gamma source for a high-energy calibration campaign. These pauses temporarily deteriorated the gas purity, but the previous level was recovered after a few days of purification. Xenon purity then reached a plateau with an electron lifetime of approximately 41~ms. This represents an improvement by a factor of $\approx$4 with respect to \NEW \cite{NEW2nubb}. Moreover, and in contrast to \NEW, in NEXT-100 we do not observe a strong dependence of the electron lifetime on $(x,y)$, and any minor variations are factored in the 3D map described in Section~\ref{sec:detector_response}. This measurement with \KR decays yields a value a factor of two lower than our measurement with alpha particles in \cite{NEXT:2025yqw}. The latter, however, was performed during the commissioning period, using an ambient-temperature (cold) gas purifier, whereas the present measurement employs a heated purifier. A subsequent electron lifetime measurement with alpha particles, using hot-getter gas purification, is consistent with the current result based on \KR decays.

The fourth panel shows the evolution of the integrated tracking-plane response. Small variations, below 1.5\%, are attributed to variations in the event distribution combined with the inhomogeneities in the tracking plane response.
Finally, the bottom panel displays the drift velocity, computed as described in Section~\ref{sec:driftv}, and shows a remarkably consistent value over time, confirming the excellent stability of the detector.

\begin{figure*}
    \centering
    \includegraphics[width=\textwidth]{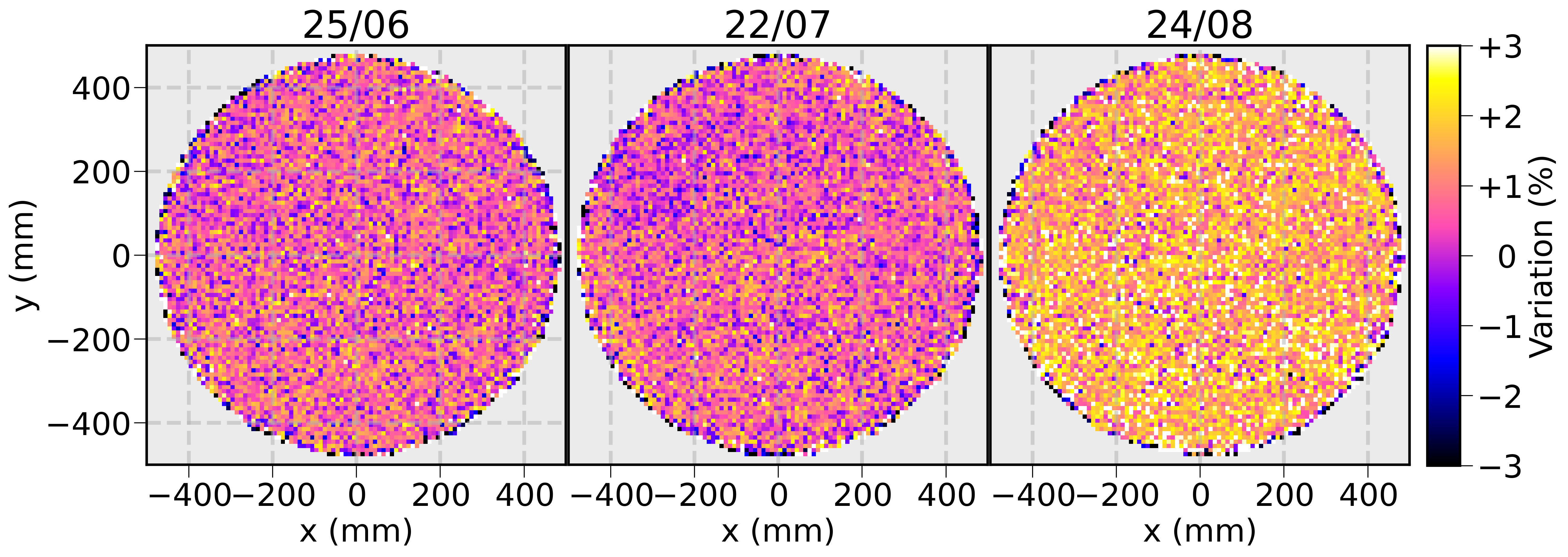}
    \caption{Variation over time in the response map with respect to a run taken on 04/06/2025. The title in each panel indicates the date of data taking for each run.}
    \label{fig:map_evolution}
\end{figure*}

Furthermore the response maps discussed in Section~\ref{sec:detector_response} are also monitored. A map is produced for each 24-hour run. \Fig~\ref{fig:map_evolution} shows the ratio between the response maps obtained for three different runs and a response map obtained at the beginning of this data-taking period (04/06/2025). These runs were taken approximately one month apart.
Small variations ($\lesssim1$\%) were observed over on the time scale of $\sim$1~month. The main contribution to these variations is the change in the overall scale of the detector response as shown in the second panel of \Fig.~\ref{fig:evolution}. Minor local (x,y,z) variations were also observed over these long time scales. This continuous \KR calibration of the detector allows to compensate for these effects and extract the maximum performance.

\section{Performance}
\label{sec:performance}



Energy resolution is one of the most critical parameters of the NEXT technology. In order to estimate it from \KR decays, raw S2 signals are corrected by the spatial variations in the detector response using the procedure described in Section~\ref{sec:calibration_procedure}. The map used in this analysis was computed with an independent data sample, from the previous run.

The corrected energy is computed using the expression

$$E = \frac{S_2}{S_0(x, y, z)} \cdot E_0, $$

\noindent
where $S_2$ is the uncorrected S2 signal in pe, $E_0 = 41.55$~keV is the known energy deposited by a \KR decay, and $S_0(x, y, z)$ is the average energy of \KR events from the corresponding voxel of the energy map shown in \Fig~\ref{fig:E0_map}. The $(x, y)$ coordinates of each decay were estimated from the barycenter of the SiPM signals, and the $z$ coordinate from the difference between S1 and S2 times, as described above.

A normal distribution of width $\sigma$ and mean $\mu$ is fitted to the energy spectrum, from which we extract the energy resolution expressed in FWHM as $R = \sqrt{8 \ln 2}\ \sigma/\mu$.

\begin{figure}
    \centering
    \includegraphics[width=\columnwidth]{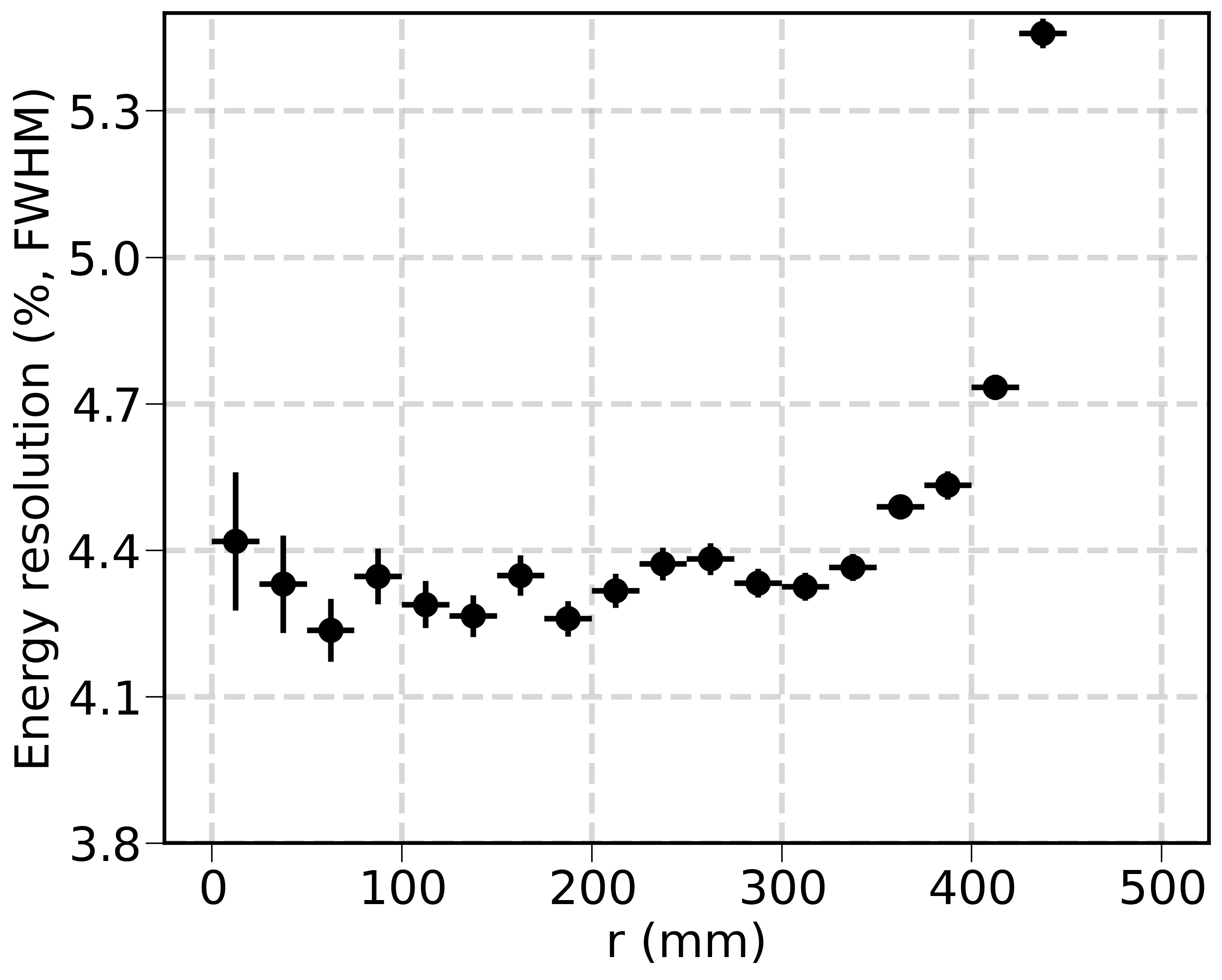}
    \caption{Energy resolution for \KR events as function of the radial position. A steep degradation is observed for $r>425$~mm.}
    \label{fig:eres_vs_r}
\end{figure}

As a result of the inhomogeneities in the detector response, the energy resolution depends on $(x,y,z)$. The dominant factor is the steep variation of the light collection efficiency close to the walls of the TPC. This is illustrated in \Fig~\ref{fig:eres_vs_r}, where we display the energy resolution as a function of the radius for the entire active volume. A clear degradation is observed above $r>425$~mm. This is attributed to the large gradient in the detector response near the edges. This combined with the increased size of the electron cloud for long drift times, introduce large fluctuations that deteriorate the energy resolution. In the subsequent analysis, only events with $r<425$~mm are considered.

\begin{figure}
    \centering
    \includegraphics[width=\columnwidth]{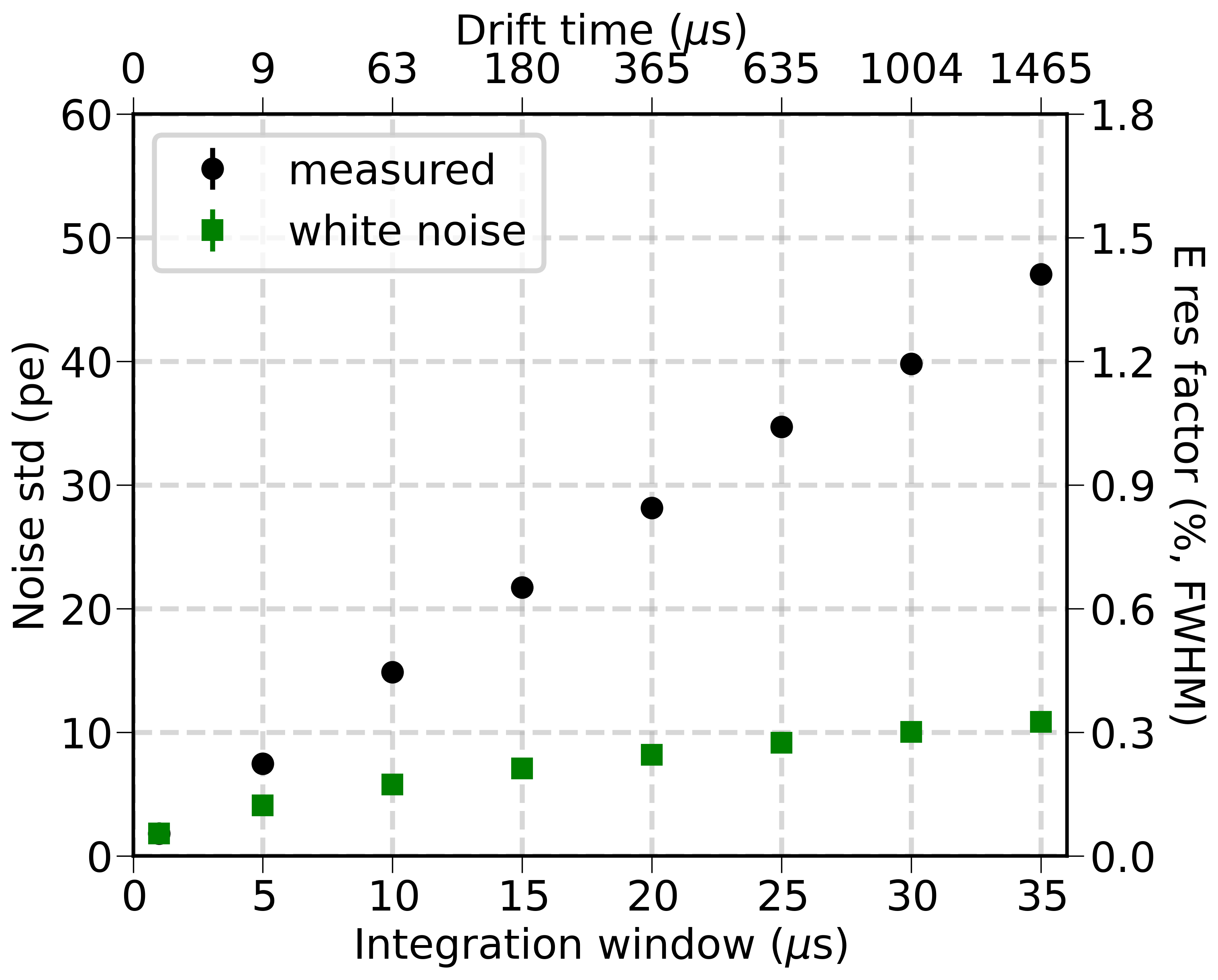}
    \includegraphics[width=\columnwidth]{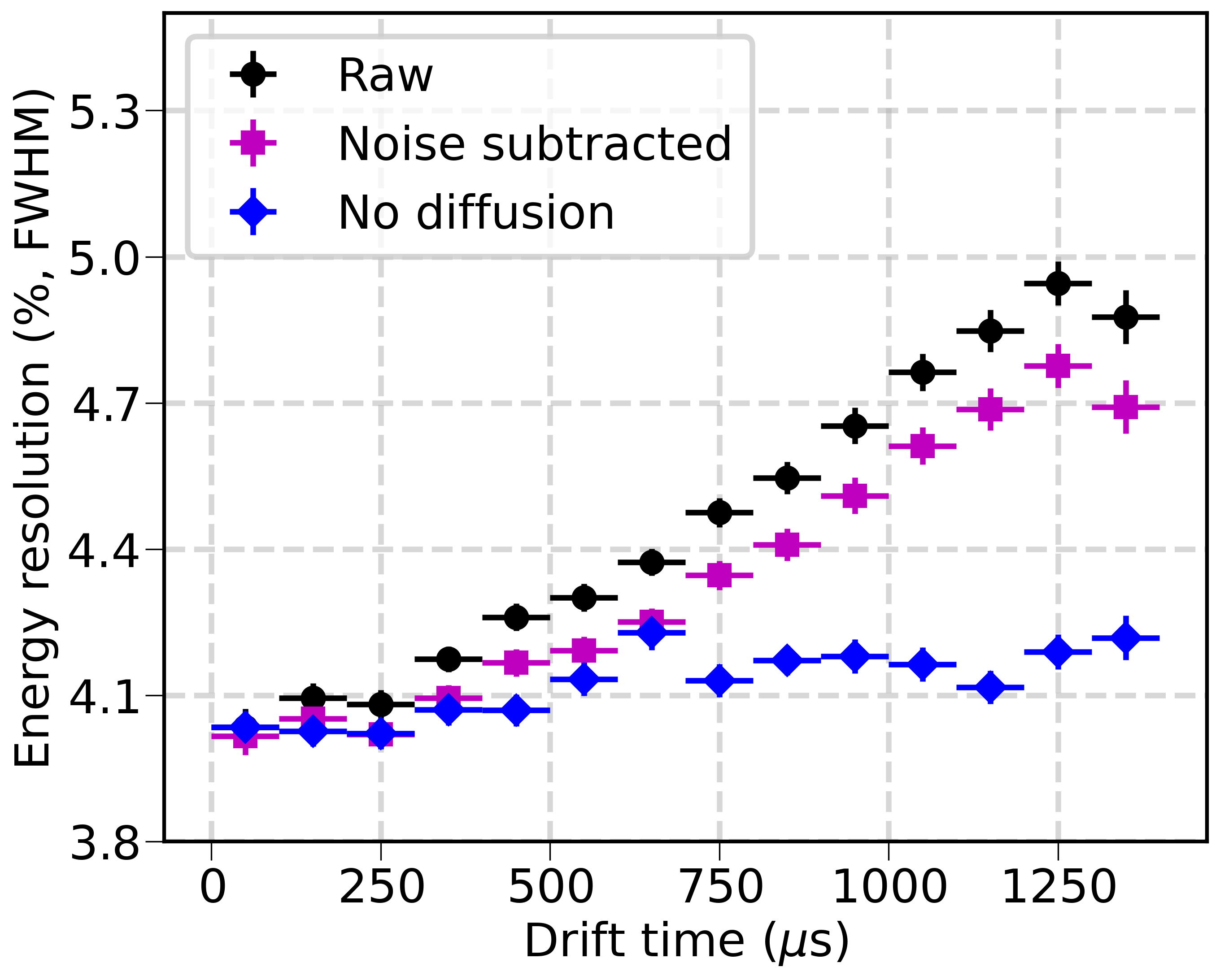}
    \caption{Top: Standard deviation of the integrated noise (black circles) as a function of the size of the integration window. The translation into drift time and its contribution to the energy resolution is displayed as a secondary x- and y-axis, respectively. The trend assuming the noise is purely white is shown as green squares.
    Bottom: Raw (black circles) energy resolution for \KR events within $r<425$~mm as a function of the drift time. The dependence of the energy resolution with drift time without the contribution of noise (magenta squares) and diffusion (blue diamonds) are also shown.}
    \label{fig:eres_vs_dt}
\end{figure}

The energy resolution also depends on the drift time (or conversely, on $z$). In this case, multiple factors contribute to its degradation. First, the transverse electron diffusion increases the size of the ionization cloud amplified in the EL region. Larger fluctuations in the number of collected photons are therefore expected, since the emission area contains regions with different light collection efficiencies. This hypothesis has been validated using our Monte Carlo simulation, using a sample with negligible electron diffusion. Second, due to the longitudinal electron diffusion, the time span of the S2 signal increases for longer drift times. The longer integration windows needed to fully measure the event energy accumulate larger noise fluctuations, worsening the energy resolution. This is illustrated in the top panel of \Fig~\ref{fig:eres_vs_dt}, where we show the standard deviation of the integrated noise (black circles) as a function of the integration window. The translation into drift time and the contribution to the energy resolution are also displayed as secondary axes. The fluctuations on the integrated noise grow faster than the square root law expected for white noise (displayed as green squares), which is attributed to a noise correlation between PMTs.
A third, smaller, contribution comes from reconstruction effects. Due to the wider shape of the pulse for longer drift times, the signal-to-noise ratio becomes smaller, leading to a less accurate peak identification, which introduces small fluctuations on the measured energy. This effect is also observed in our simulations.

Note that, of these contributions, the first and the third ones will be significantly reduced at the nominal operating pressure of \NEXT (13.5 bar), while the second becomes negligible at higher energies, where the physics program of the NEXT experiment is focused. 

The bottom panel of \Fig~\ref{fig:eres_vs_dt} shows the energy resolution for \KR events at $r<425$~mm  as a function of the drift time. The black circles correspond to the raw values, where we observe a linear degradation as the drift time increases. The magenta squares show the energy resolution with the noise contribution subtracted and the blue diamonds showcase the scenario with no noise and no electron diffusion based on the trend from our Monte Carlo simulation. These two dominant contributions account almost entirely for the dependence of the energy resolution with drift time.

This run of the NEXT-100 detector at reduced pressure is not fully representative of the standard operating conditions expected for the upcoming physics data-taking run. Under realistic conditions --- 13.5 bar pressure and a drift field of 200-400 V/cm --- the electron diffusion coefficient is expected to decrease by roughly a factor two. This reduction allows for the definition of a fiducial volume with $r<425$ mm and $z<605$ mm, consistent with the detector’s anticipated operating regime in the near future.

\begin{figure}
    \centering
    \includegraphics[width=\columnwidth]{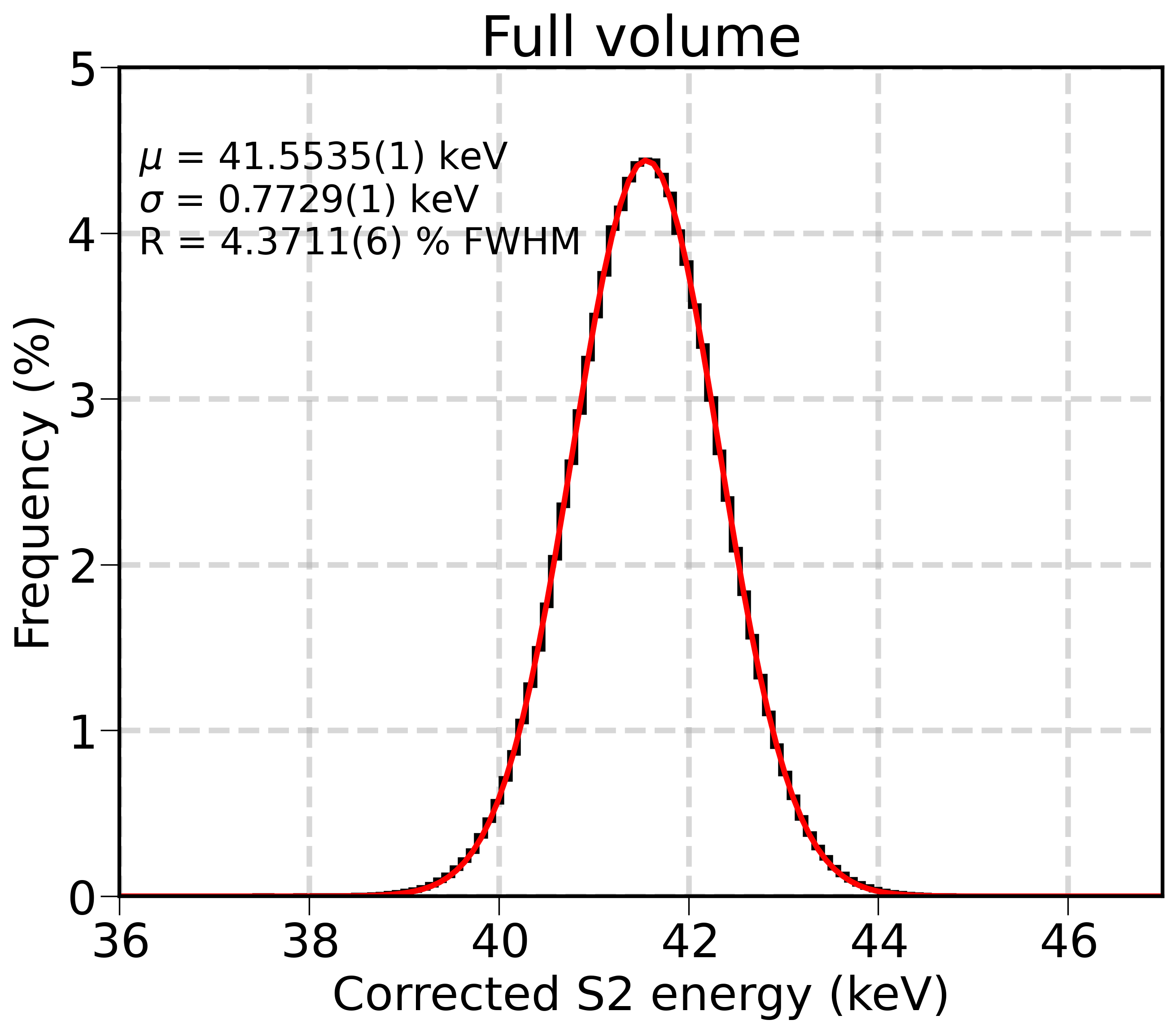}
    \includegraphics[width=\columnwidth]{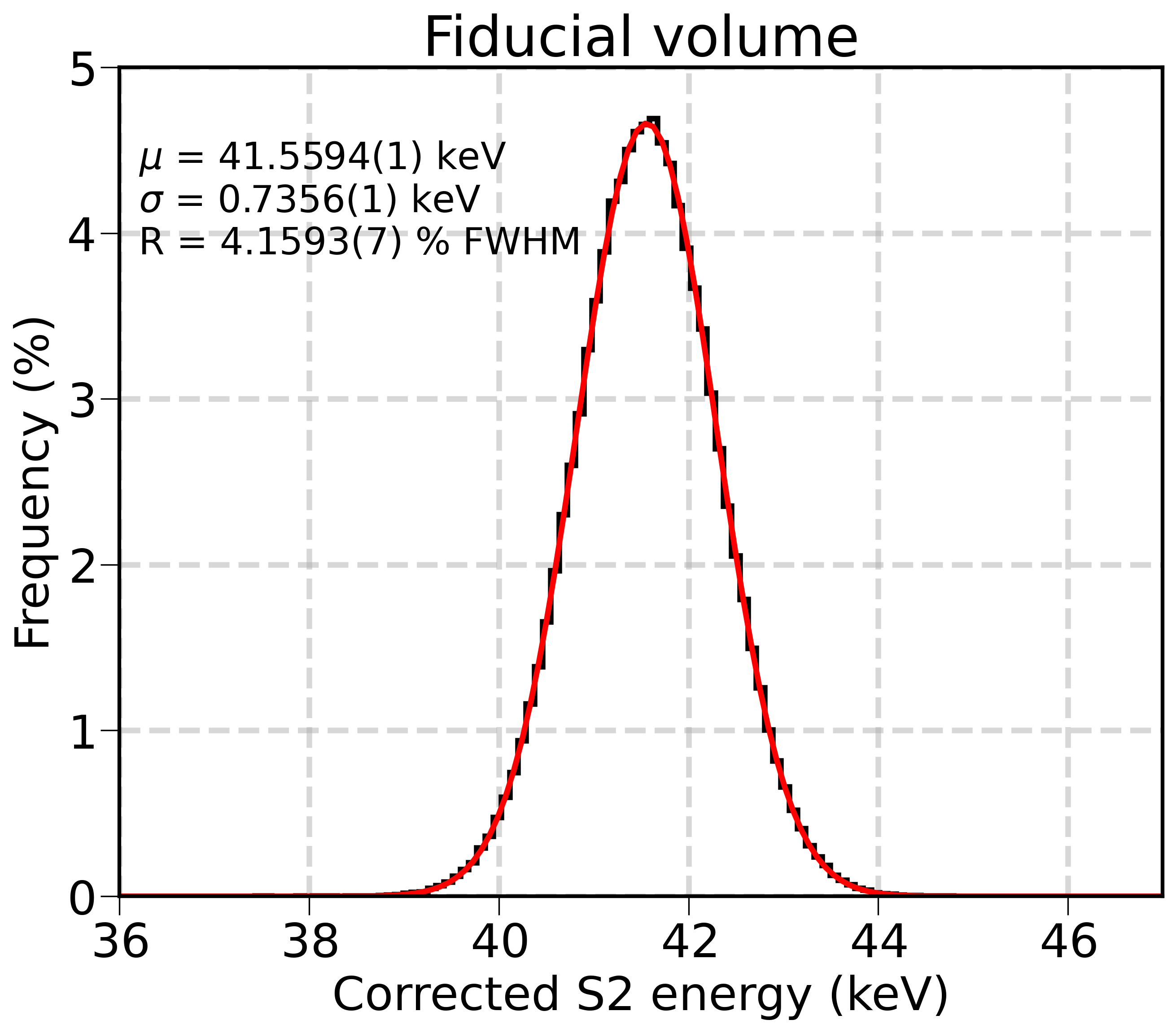}
    \caption{Energy spectrum after corrections of selected \KR events in the full volume (top) and in a fiducial volume (bottom) defined by $r<425$~mm and $z<605$~mm. A Gaussian distribution, shown as a red solid line, is fitted to the data yielding energy resolutions at 41.5 keV of 4.37\% FWHM and 4.16\% FWHM in the full and fiducial volumes, respectively.}
    \label{fig:E_resolution}
\end{figure}

The top panel of \Fig~\ref{fig:E_resolution} shows the corrected energy spectrum of \KR events at $r<425$~mm. Superimposed on the histogram is a fit to a Gaussian function, from which an energy resolution of 4.37\% FWHM at 41.5~keV is extracted. In the aforementioned fiducial volume ($r<425$~mm and $z<605$~mm) the energy resolution improves to 4.16\%. The statistical uncertainty in these measurements is negligible. This result  represents a slight improvement from the one previously obtained in \NEW using \KR\ calibration data (4.55 \% FWHM in the full volume) \cite{NEWKR}.

Assuming the energy resolution is dominated by photon-counting statistics, it is expected to scale with energy as $1/\sqrt{E}$. As demonstrated in \NEW, the reconstruction for extended tracks introduces other terms to the energy resolution, worsening it slightly \cite{NEWEres2}. However, we take this extrapolation as a benchmark of the detector performance at high energies. Following this extrapolation, the resolution at the \Qbb value (2.458~MeV) improves by a factor $\sqrt{E_0/Q_{\beta\beta}}$. Thus, the extrapolated energy resolutions at \Qbb are 0.57\%~FWHM in the full detector volume and 0.54\% FWHM in the fiducial region.
These values are well below the sub-percent FWHM resolution target set in the detector design.

\section{Conclusions}
\label{sec:conclusions}

We presented the first results obtained with the \NEXT\ detector operating at 4~bar using \KR calibration data. Using the high-statistics \KR sample, we mapped the detector response over the entire active volume, which are a combination of variations in the light collection efficiency and losses due to electron attachment. From this, we derived 3-dimensional correction factors to homogenize the S2 detector response. After applying these corrections to \KR data, we achieve an energy resolution at 41.5~keV of 4.37\% FWHM in the full detector volume. We observe a degradation of the energy resolution with radial and longitudinal positions. Both of these effects are attributed in part to the high electron diffusion in the current conditions. This led us to define a fiducial volume constrained to $r<425$~mm and $z<605$~mm, which is representative of the nominal conditions of the upcoming physics run at 13.5~bar. In this volume, the energy resolution improves to 4.16\% FWHM. Extrapolating these results to the \Qbb energy, we estimate an energy resolution close to 0.5\% FWHM.

Furthermore, the detector has demonstrated excellent stability during the initial data-taking period with the PMT response varying $\lesssim$2\% over a three-month period, and a stable electron lifetime of $\sim$41~ms, much longer than the maximum drift time.

These results validate the performance of the \NEXT\ detector under initial operating conditions and reaffirm the viability and stability of HPXeTPC-EL detectors for the search for neutrinoless double beta decay.

\begin{acknowledgements}
The NEXT Collaboration acknowledges support from the following agencies and institutions: the European Research Council (ERC) under Grant Agreemenst No. 951281-BOLD and\linebreak 101039048-GanESS; the European Union’s Framework Programme for Research and Innovation Horizon 2020 (2014–2020) under Grant Agreement No. 860881-HIDDeN; the MCIN/AEI of Spain and ERDF A way of making Europe under grants PID2021-125475NB and RTI2018-095979, and the Severo Ochoa and Mar\'ia de Maeztu Program grants CEX2023-001292-S, CEX2023-001318-M and CEX2018-000867-S; the Generalitat Valenciana of Spain under grants PROMETEO/2021/087, ASFAE/2022/028, ASFAE/2022/029, CISEJI/2023/27 and \linebreak CIDEXG/2023/16; the Department of Education of the Basque Government of Spain under the predoctoral training program non-doctoral research personnel; the Spanish la Caixa Foundation (ID 100010434) under fellowship code LCF/BQ/PI22/11910019; the Portuguese FCT under project UID/FIS/04559/2020 to fund the activities of LIBPhys-UC; the Israel Science Foundation (ISF) under grant 1223/21; the Pazy Foundation (Israel) under grants 310/22, 315/19 and 465; the US Department of Energy under contracts number DE-AC02-06CH11357 (Argonne National Laboratory), DE-AC02-07CH11359 (Fermi National Accelerator Laboratory), DE-FG02-13ER42020 (Texas A\&M), DE-SC0019054 (Texas Arlington) and DE-SC0019223 (Texas Arlington); the US National Science Foundation under award number NSF CHE 2004111; the Robert A Welch Foundation under award number Y-2031-20200401. Finally, we are grateful to the Laboratorio Subterr\'aneo de Canfranc for hosting and supporting the NEXT experiment.
\end{acknowledgements}

\bibliographystyle{unsrturl}
\bibliography{pool/bibliography}

\begin{thebibliography}{10}

\bibitem{KamLAND-Zen:23}
S.~Abe et~al.
\newblock {Search for the Majorana Nature of Neutrinos in the Inverted Mass Ordering Region with KamLAND-Zen}.
\newblock {\em Physical Review Letters}, 130(5), January 2023.
\newblock URL: \url{http://dx.doi.org/10.1103/PhysRevLett.130.051801}, \href {https://doi.org/10.1103/physrevlett.130.051801} {\path{doi:10.1103/physrevlett.130.051801}}.

\bibitem{25tk-nctn}
H.~Acharya~et al.
\newblock First results on the search for lepton number violating neutrinoless double-$\ensuremath{\beta}$ decay with the legend-200 experiment.
\newblock {\em Phys. Rev. Lett.}, pages~--, Sep 2025.
\newblock URL: \url{https://link.aps.org/doi/10.1103/25tk-nctn}, \href {https://doi.org/10.1103/25tk-nctn} {\path{doi:10.1103/25tk-nctn}}.

\bibitem{DEMO1}
V~\'Alvarez et~al.
\newblock Initial results of next-demo, a large-scale prototype of the next-100 experiment.
\newblock {\em Journal of Instrumentation}, 8(04):P04002, apr 2013.
\newblock URL: \url{https://dx.doi.org/10.1088/1748-0221/8/04/P04002}, \href {https://doi.org/10.1088/1748-0221/8/04/P04002} {\path{doi:10.1088/1748-0221/8/04/P04002}}.

\bibitem{DEMO2}
V.~\'Alvarez et~al.
\newblock {Operation and first results of the NEXT-DEMO prototype using a silicon photomultiplier tracking array}.
\newblock {\em JINST}, 8:P09011, 2013.
\newblock \href {https://doi.org/10.1088/1748-0221/8/09/P09011} {\path{doi:10.1088/1748-0221/8/09/P09011}}.

\bibitem{DEMO3}
V.~\'Alvarez et~al.
\newblock {Near-Intrinsic Energy Resolution for 30 to 662 keV Gamma Rays in a High Pressure Xenon Electroluminescent TPC}.
\newblock {\em Nucl. Instrum. Meth.}, A708:101–114, 2012.
\newblock URL: \url{https://doi.org/10.1016/j.nima.2012.12.123}.

\bibitem{DEMO4}
P.~Ferrario et~al.
\newblock {First proof of topological signature in the high pressure xenon gas TPC with electroluminescence amplification for the NEXT experiment}.
\newblock {\em JHEP}, 2016:104, 2016.
\newblock URL: \url{https://doi.org/10.1007/JHEP01(2016)104}.

\bibitem{NEWDetector}
F.~Monrabal et~al.
\newblock {The NEXT White (NEW) detector}.
\newblock {\em JINST}, 13:P12010, 2018.
\newblock URL: \url{https://doi.org/10.1088/1748-0221/13/12/P12010}.

\bibitem{NEWEres1}
J.~Renner et~al.
\newblock {Initial results on energy resolution of the NEXT-White detector}.
\newblock {\em JINST}, 13:P10020, 2018.
\newblock URL: \url{https://doi.org/10.1088/1748-0221/13/12/P12010}.

\bibitem{NEWEres2}
J.~Renner et~al.
\newblock {Energy calibration of the NEXT-White detector with 1\% resolution near $Q_{\beta\beta}$ of $^{136}$Xe}.
\newblock {\em JHEP}, 10:230, 2019.
\newblock URL: \url{https://doi.org/10.1007/JHEP10%282019%29230}.

\bibitem{Kekic2021}
M.~Kekic et~al.
\newblock {Demonstration of background rejection using deep convolutional neural networks in the NEXT experiment}.
\newblock {\em Journal of High Energy Physics}, 2021(1):189, 2021.
\newblock \href {https://doi.org/10.1007/JHEP01(2021)189} {\path{doi:10.1007/JHEP01(2021)189}}.

\bibitem{NEWTrack1}
P.~Ferrario et~al.
\newblock {Demonstration of the event identification capabilities of the NEXT-White detector}.
\newblock {\em JHEP}, 10:052, 2019.
\newblock URL: \url{https://doi.org/10.1007/JHEP10%282019%29052}.

\bibitem{NEWTrack2}
A.~Sim\'on et~al.
\newblock {Boosting background suppression in the NEXT experiment through Richardson-Lucy deconvolution}.
\newblock {\em JHEP}, 21:146, 2020.
\newblock URL: \url{https://doi.org/10.1007/JHEP07%282021%29146}.

\bibitem{NEWBkg1}
P.~Novella et~al.
\newblock {Measurement of radon-induced backgrounds in the NEXT double beta decay experiment}.
\newblock {\em JHEP}, 10:112, 2018.
\newblock URL: \url{https://doi.org/10.48550/arXiv.1804.00471}.

\bibitem{NEWBkg2}
P.~Novella et~al.
\newblock {Radiogenic Backgrounds in the NEXT Double Beta Decay Experiment}.
\newblock {\em JHEP}, 10:051, 2019.
\newblock URL: \url{https://doi.org/10.1007/JHEP10%282019%29051}.

\bibitem{NEW2nubb}
P.~Novella et~al.
\newblock {Measurement of the $^{136}$Xe two-neutrino double $\beta$ decay half-life via direct background subtraction in NEXT}.
\newblock {\em Phys. Rev. C}, 105:055501, 2022.
\newblock URL: \url{https://doi.org/10.1103/PhysRevC.105.055501}.

\bibitem{NEW0nubb}
P.~Novella et~al.
\newblock {Demonstration of neutrinoless double beta decay searches in gaseous xenon with NEXT}.
\newblock {\em JHEP}, 09:190, 2023.
\newblock URL: \url{https://doi.org/10.1007/JHEP09%282023%29190}.

\bibitem{NEXT:2025yqw}
C.~Adams et~al.
\newblock {The NEXT-100 Detector}.
\newblock 5 2025.
\newblock Approved at EPJC.
\newblock \href {https://arxiv.org/abs/2505.17848} {\path{arXiv:2505.17848}}.

\bibitem{NEWKR}
G.~Martínez-Lema et~al.
\newblock {Calibration of the NEXT-White detector using $^{83m}$Kr decays}.
\newblock {\em Journal of Instrumentation}, 13(10):P10014–P10014, October 2018.
\newblock URL: \url{http://dx.doi.org/10.1088/1748-0221/13/10/P10014}, \href {https://doi.org/10.1088/1748-0221/13/10/p10014} {\path{doi:10.1088/1748-0221/13/10/p10014}}.

\bibitem{KrSource}
D.~Vénos, A.~Špalek, O.~Lebeda, and M.~Fišer.
\newblock {$^{83m}$Kr radioactive source based on $^{83}$Rb trapped in cation-exchange paper or in zeolite}.
\newblock {\em Applied Radiation and Isotopes}, 63(3):323--327, 2005.
\newblock URL: \url{https://www.sciencedirect.com/science/article/pii/S0969804305001260}, \href {https://doi.org/10.1016/j.apradiso.2005.04.011} {\path{doi:10.1016/j.apradiso.2005.04.011}}.

\bibitem{PMTelectronics}
V.~Álvarez, V.~Herrero-Bosch, R.~Esteve, A.~Laing, J.~Rodríguez, M.~Querol, F.~Monrabal, J.F. Toledo, and J.J. Gómez-Cadenas.
\newblock {The electronics of the energy plane of the NEXT-White detector}.
\newblock {\em Nuclear Instruments and Methods in Physics Research Section A: Accelerators, Spectrometers, Detectors and Associated Equipment}, 917:68--76, 2019.
\newblock URL: \url{https://www.sciencedirect.com/science/article/pii/S0168900218317790}, \href {https://doi.org/10.1016/j.nima.2018.11.126} {\path{doi:10.1016/j.nima.2018.11.126}}.

\bibitem{NEXT:2023jwt}
J.~Haefner et~al.
\newblock {Demonstration of event position reconstruction based on diffusion in the NEXT-white detector}.
\newblock {\em Eur. Phys. J. C}, 84(5):518, 2024.
\newblock \href {https://arxiv.org/abs/2311.03441} {\path{arXiv:2311.03441}}, \href {https://doi.org/10.1140/epjc/s10052-024-12865-9} {\path{doi:10.1140/epjc/s10052-024-12865-9}}.

\bibitem{NEWdif}
A.~Simón et~al.
\newblock {Electron drift properties in high pressure gaseous xenon}.
\newblock {\em Journal of Instrumentation}, 13(07):P07013–P07013, July 2018.
\newblock URL: \url{http://dx.doi.org/10.1088/1748-0221/13/07/P07013}, \href {https://doi.org/10.1088/1748-0221/13/07/p07013} {\path{doi:10.1088/1748-0221/13/07/p07013}}.

\bibitem{MAGBOLTZ}
S.F. Biagi.
\newblock {Monte Carlo simulation of electron drift and diffusion in counting gases under the influence of electric and magnetic fields}.
\newblock {\em Nuclear Instruments and Methods in Physics Research Section A: Accelerators, Spectrometers, Detectors and Associated Equipment}, 421(1):234--240, 1999.
\newblock URL: \url{https://www.sciencedirect.com/science/article/pii/S0168900298012339}, \href {https://doi.org/10.1016/S0168-9002(98)01233-9} {\path{doi:10.1016/S0168-9002(98)01233-9}}.

\bibitem{DEMO_2013}
V~\'Alvarez et~al.
\newblock Ionization and scintillation response of high-pressure xenon gas to alpha particles.
\newblock {\em Journal of Instrumentation}, 8(05):P05025, may 2013.
\newblock URL: \url{https://dx.doi.org/10.1088/1748-0221/8/05/P05025}, \href {https://doi.org/10.1088/1748-0221/8/05/P05025} {\path{doi:10.1088/1748-0221/8/05/P05025}}.

\bibitem{Lorca_2014}
D~Lorca et~al.
\newblock {Characterisation of NEXT-DEMO using xenon K$\alpha$ X-rays}.
\newblock {\em Journal of Instrumentation}, 9(10):P10007, oct 2014.
\newblock URL: \url{https://dx.doi.org/10.1088/1748-0221/9/10/P10007}, \href {https://doi.org/10.1088/1748-0221/9/10/P10007} {\path{doi:10.1088/1748-0221/9/10/P10007}}.

\end{thebibliography}
\end{document}